\documentclass[11pt]{article}
\usepackage{moriond}
\usepackage{amsmath}
\usepackage{amssymb}
\newcommand{\Photo}{}

\begin{document}
\vspace*{4cm}
\title{Fitting the Two-Higgs-Doublet model of type II}
\author{Otto Eberhardt}
\address{Istituto Nazionale di Fisica, Sezione di Roma,\\
Piazzale Aldo Moro 2, I-00185 Rome, Italy}

\maketitle\abstracts{
We present the current status of the Two-Higgs-Doublet model of type II. Taking into account all available relevant information, we exclude at $95$\% CL sizeable deviations of the so-called alignment limit, in which all couplings of the light CP-even Higgs boson $h$ are Standard-Model-like. While we can set a lower limit of $240$ GeV on the mass of the pseudoscalar Higgs boson at $95$\% CL, the mass of the heavy CP-even Higgs boson $H$ can be even lighter than $200$ GeV. The strong constraints on the model parameters also set limits on the triple Higgs couplings: the $hhh$ coupling in the Two-Higgs-Doublet model of type II cannot be larger than in the Standard Model, while the $hhH$ coupling can maximally be $2.5$ times the size of the Standard Model $hhh$ coupling, assuming an $H$ mass below $1$ TeV. The selection of benchmark scenarios which maximize specific effects within the allowed regions for further collider studies is illustrated for the $H$ branching fraction to fermions and gauge bosons. As an example, we calculate the cross section of $gg\to hh$ for four benchmark points and show that a resonant $H$ could enhance it by almost a factor of $70$ at a centre-of-mass energy of $14$ TeV.}

\section{Introduction}

LHC measurements of the first run have revealed the existence of a scalar particle at $126$ GeV and determined its couplings to the known existing particles of the Standard Model (SM) with an astonishing precision \cite{Aad:2012tfa,Chatrchyan:2012ufa}. Even if none of these experimental results stirs doubts about this new particle being the scalar boson predicted by the BEH mechanism \cite{Higgs:1964pj,Englert:1964et} in the framework of the SM, it could be that the scalar sector contains more than one particle generation \cite{Lee:1973iz}, just as the fermionic sector. An appealing non-minimal model is the Two-Higgs-Doublet model of type II (2HDM$I\!I$) \cite{Gunion:2002zf,Branco:2011iw}. Several studies in the recent years have shown that the parameter space of this model is strongly constrained by various theoretical and experimental bounds. In these proceedings, we present the current status of the 2HDM$I\!I$ with a CP conserving scalar potential and a soft $Z_2$ symmetry breaking term, combining Higgs measurements with electroweak precision observables and the relevant flavour constraints in our analysis. With the result of the global fit at hand, we study the possible size of triple-Higgs couplings to lay the ground for analyses of Higgs pair production cross sections at the LHC. We quantify the possible enhancement factor of the $hh$ production cross section at the 14 TeV run of the LHC.
All results can be found in our last publications \cite{Eberhardt:2013uba,Baglio:2014nea}, where the interested reader can find the details about the constraints and the fitting procedure.

\section{Fit constraints}

Theoretical considerations enable us to set limits for the allowed values of the quartic Higgs couplings: Apart from the requirement of an absolute minimum of the scalar potential at around $246$ GeV, we apply the upper limit of $1/8$ for the absolute value of the eigenvalues of the tree-level $S$ matrix, because we want the Higgs self-couplings to stay perturbative. This has been proven to be a reliable upper bound for the SM quartic coupling \cite{Nierste:1995zx}. (For a comparison with the more conservative bound of 1, we refer to our latest article \cite{Baglio:2014nea}.)
In order to discuss the experimental constraints, we want to use the physical parameters, which consist of the four masses $m_h$, $m_H$, $m_A$ and $m_{H^+}$, the two diagonalisation angle combinations $\tan \beta$ and $\beta-\alpha$, the vacuum expectation value $v$ as well as the soft $Z_2$ breaking parameter $m_{12}^2$. From all available experimental data, we combined the following measurements: electroweak precision observables, the branching ratio of radiative $B$ meson decays $BR(B\to X_s \gamma)$, the mass difference of the $B_s$ meson and LHC light and heavy Higgs search results. For the first time we also added the exclusion limits to the $gg\to H\to hh$ and $gg\to A \to hZ$ cross sections \cite{CMS-PAS-HIG-13-025} to our set of experimental inputs. However, those two observables do not have any effect of the results presented in the latest publication \cite{Baglio:2014nea}. While the figures in the mentioned article stem from fits with the myFitter framework \cite{Wiebusch:2012en}, the fit results shown in these proceedings were obtained with the CKMfitter package \cite{Hocker:2001xe}, thus being an independent cross-check.

\section{Results}

\begin{figure}
\begin{minipage}{0.5\linewidth}
\centerline{\includegraphics[width=0.86\linewidth,]{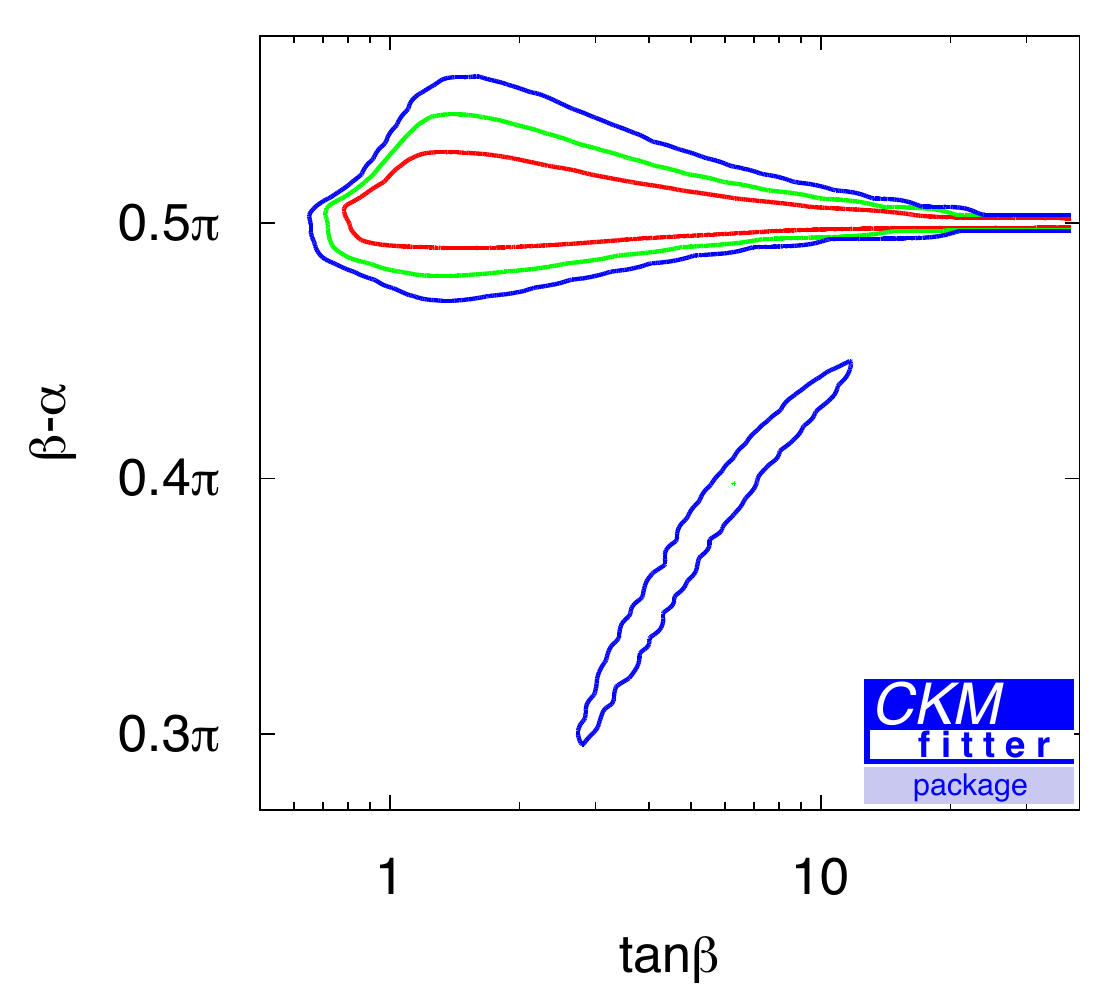}}
\end{minipage}
\hfill
\begin{minipage}{0.5\linewidth}
\centerline{\includegraphics[width=0.9\linewidth]{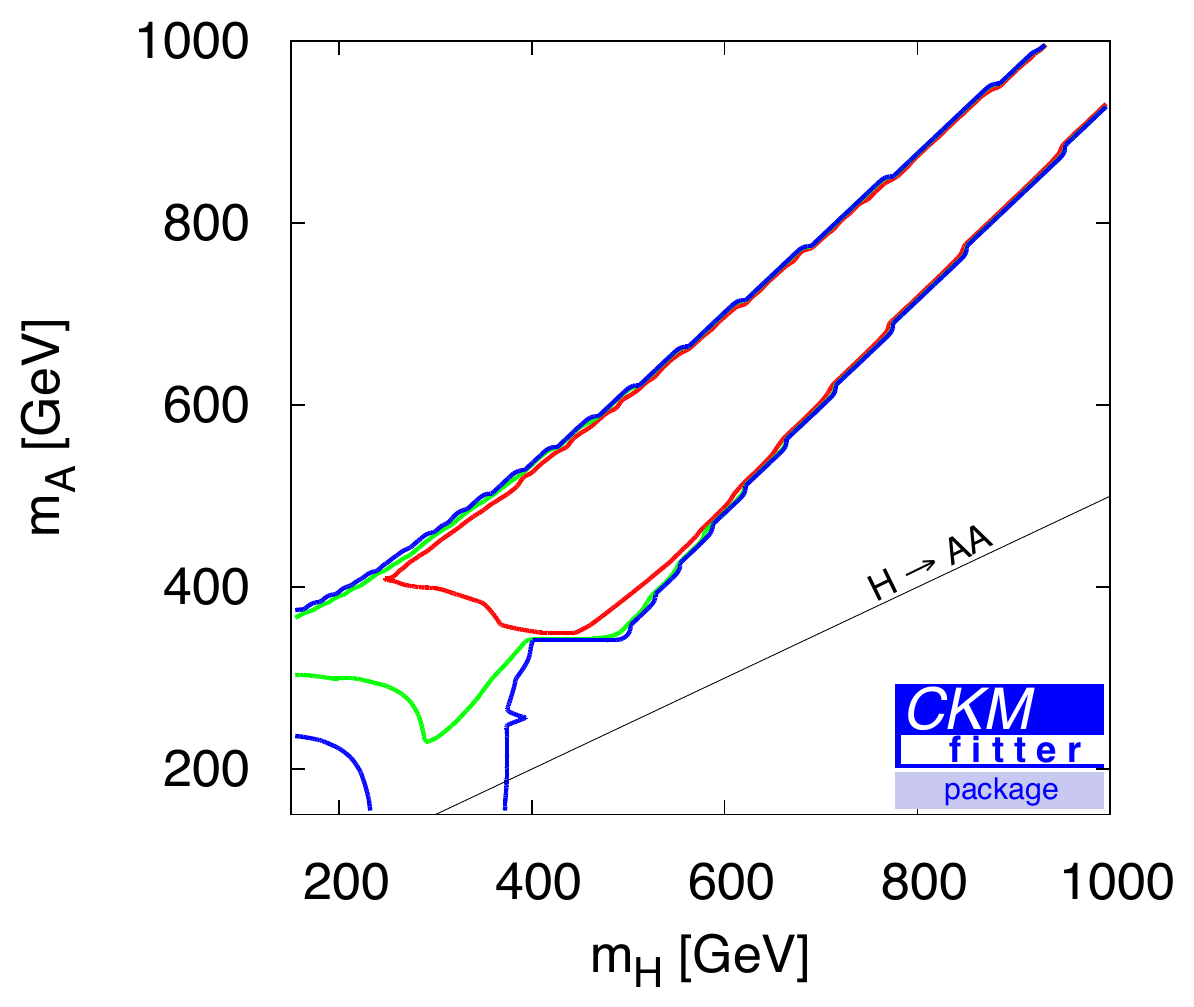}}
\end{minipage}
\caption{$1$, $2$ and $3\:\sigma$ allowed regions in the $\tan \beta$--$(\beta-\alpha)$ plane (left) and in the $m_H$--$m_A$ plane (right).}
\label{fig:fits}
\end{figure}

The scan over the $\tan \beta$-($\beta-\alpha$) plane (left side of Fig.~\ref{fig:fits}) shows that at $2\:\sigma$ $\beta-\alpha$ is now constrained to be closer as $0.05 \pi$ to the so-called alignment limit $\beta-\alpha=\pi/2$, in which $H$ does not couple to vector bosons at tree-level and all $h$ couplings to SM fermions and vector bosons are SM-like. The right side of Fig.~\ref{fig:fits} displays the allowed combinations of the $H$ and $A$ masses. At $2\:\sigma$, $m_A$ must be larger than $240$ GeV, while $m_H$ can be even lighter than $200$ GeV; the decay $H\to AA$ can be excluded.

\begin{figure}[htbp]
\begin{minipage}{0.5\linewidth}
\centering
\includegraphics[width=0.55\linewidth,]{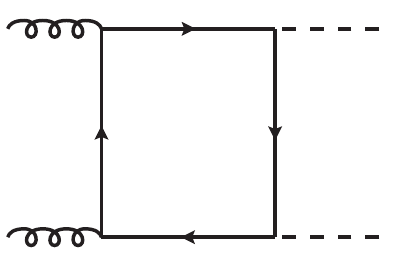}\\[10pt]
\includegraphics[width=0.55\linewidth,]{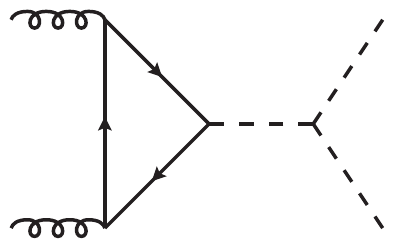}
\end{minipage}
\hfill
\begin{minipage}{0.5\linewidth}
\centerline{\includegraphics[width=0.85\linewidth]{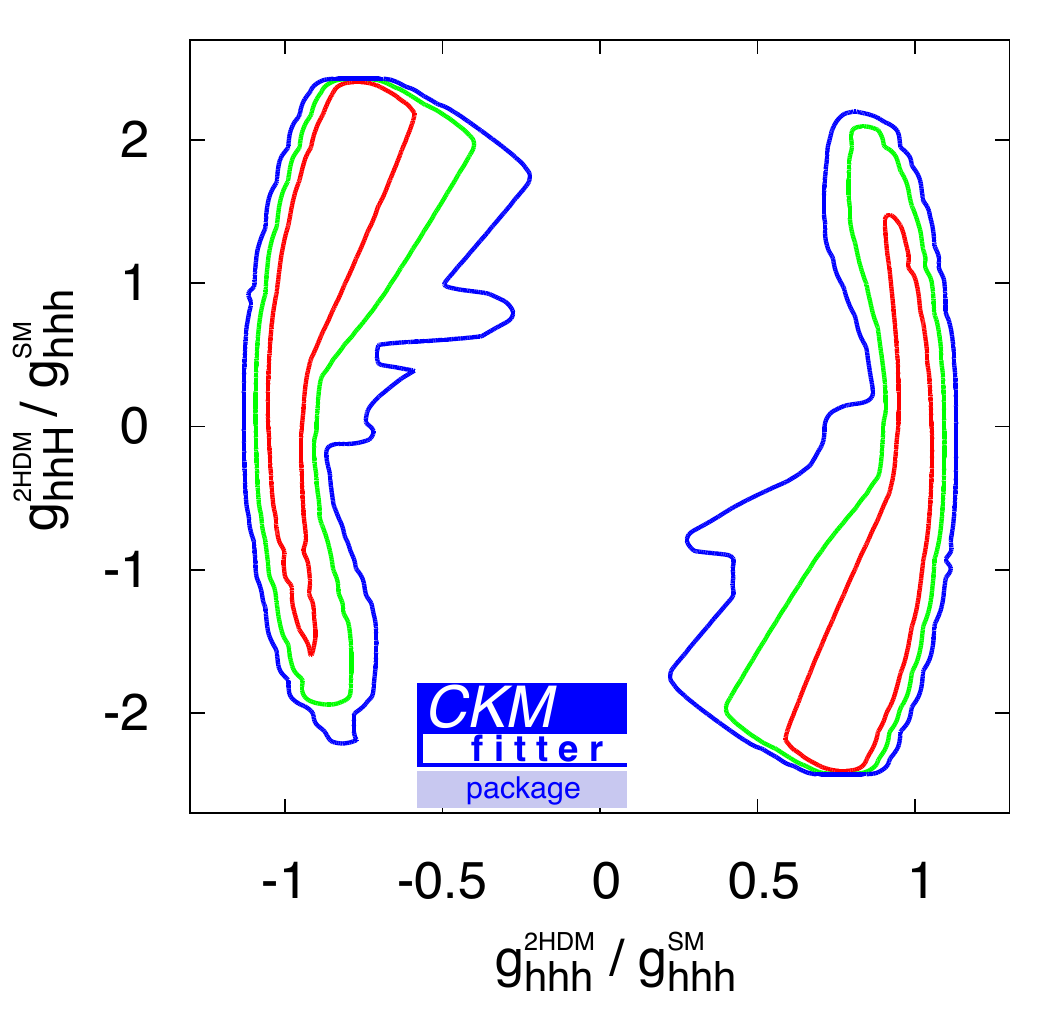}}
\end{minipage}
\caption{On the left, the relevant diagrams for the process $gg\to hh$ are given. The right figure shows the $1$, $2$ and $3\:\sigma$ allowed coupling strength of an $hhh$ vertex normalized to the SM value plotted against the same quantity for the $hhH$ vertex.}
\label{fig:hhH}
\end{figure}

If the heavy Higgs bosons are light enough for direct detection but escape the ``standard'' searches at LHC due to their coupling behaviour, they might show up in other observables. One possibility is an enhancement of the cross section of double $h$ production, which at the LHC would be mainly the process $gg\to hh$, for which the two diagrams on the left of Fig.~\ref{fig:hhH} are relevant. In the lower (``triangle'') diagram, the intermediate scalar could be an $h$ or $H$ boson in the 2HDM$I\!I$. Therefore, we need to investigate the corresponding couplings of the $hhh$ and $hhH$ vertices, $g_{hhh}$ and $g_{hhH}$.
Our analysis shows that an enhancement of the ratio $g_{hhh}/g_{hhh}^{SM}$ is not possible in the 2HDM$I\!I$, see the right side of Fig.~\ref{fig:hhH}, but it could be reduced to less than $50$\% at the $2\:\sigma$ level. The coupling $g_{hhH}$, however, can be more than twice as large as $g_{hhh}^{SM}$. (Here, we assumed $m_H\leq 1$ TeV.)

\begin{figure}
\begin{minipage}{0.5\linewidth}
\centerline{\includegraphics[width=0.9\linewidth]{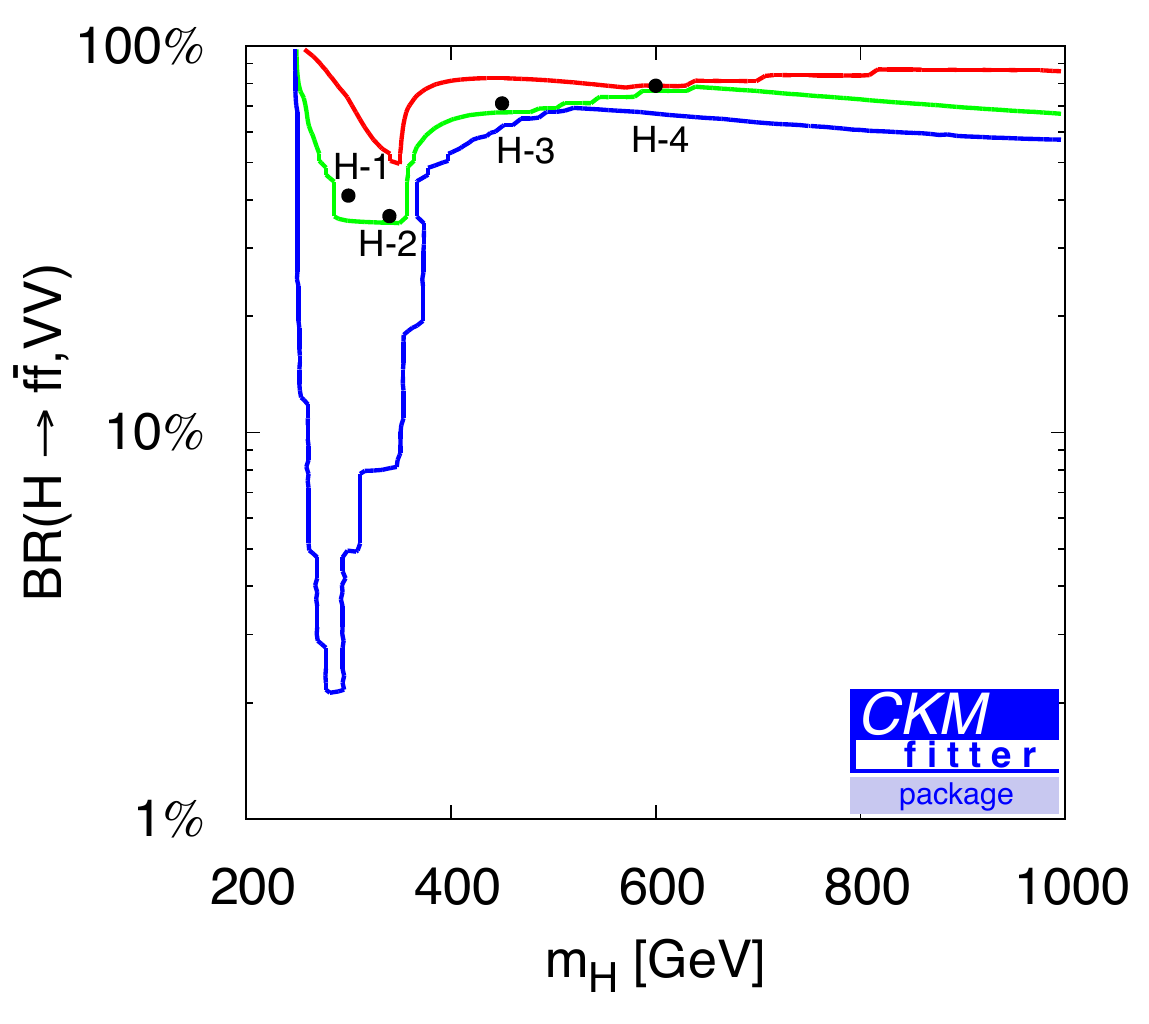}}
\end{minipage}
\hfill
\begin{minipage}{0.5\linewidth}
\centerline{\includegraphics[width=0.9\linewidth]{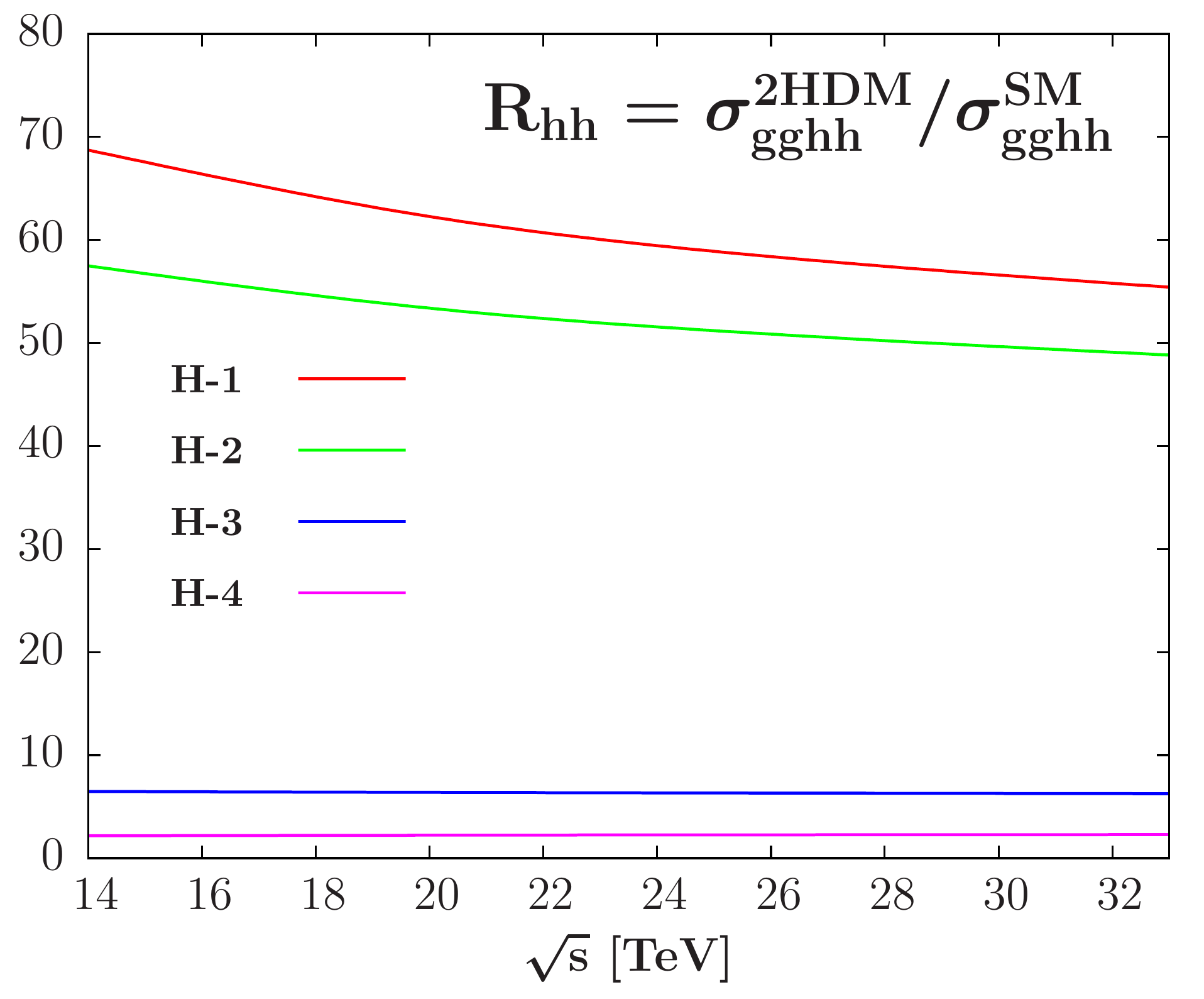}}
\end{minipage}
\caption{The figure on the left shows the $1$, $2$ and $3\:\sigma$ regions for the allowed branching ratio of an $H$ boson to SM fermions and vector bosons depending on $m_H$. The four benchmark points H-1, H-2, H-3 and H-4 feature maximal deviation from ``standard decays'' at the $95$\% level. The enhancement of the $gg\to hh$ cross section with respect to the SM value for the four chosen benchmark points is shown on the right side.}
\label{fig:gghh}
\end{figure}

However, also the branching ratio of the $H$ boson plays a crucial role: from the left of Fig.~\ref{fig:gghh}, we can see that the branching ratio of $H$ decaying into fermions and vector bosons $BR(H\to f\bar f,VV)$ could be suppressed to less than $40$\% at the $2\:\sigma$ level, depending on the $H$ mass. For a precise study of the $gg\to hh$ cross section, we define for different $H$ masses the four benchmark scenarios H-1 to H-4, in which effects of ``non-standard'' $H$ decays are maximized. On the right of Fig.~\ref{fig:gghh}, we show that for the benchmark point H-1, a resonant $H$ in the lower right diagram of Fig.~\ref{fig:hhH} could enhance the $gg\to hh$ cross section to a factor of almost $70$ with respect to the SM value, which should be visible at the next run of the LHC at $14$ TeV.

Furthermore, we studied the other 2HDM$I\!I$ triple scalar couplings $hHH$, $HHH$, $hAA$ and $hH^+H^-$, and found their maximal possible coupling strength to be limited to $5.5 g_{hhh}^{SM}$. Also the branching ratio of ``standard'' decays of $A$ bosons was analyzed. It strongly depends on the $A$ mass, and if the decay $A\to HZ$ is kinematically possible, it can be suppressed to $1$\% at the $2\:\sigma$ level if $m_A$ is just below the $t\bar t$ threshold. (Similarly, $BR(H\to f\bar f,VV)$ is minimal for $m_H\lesssim 2m_t$.)
In order to trigger more intricate collider studies, we specified a set of benchmark points which feature the largest effects currently allowed by all mentioned constraints at the $2\:\sigma$ level, varying the corresponding relevant heavy Higgs mass \cite{Baglio:2014nea}: Like H-1 to H-4 for minimal $BR(H\to f\bar f,VV)$, we provide benchmark scenarios for minimal ``standard'' $A$ decays, minimal $hhh$ as well as maximal $hhH$, $hHH$, $HHH$, $hAA$ and $hH^+H^-$ couplings.

\section{Conclusions}

We have discussed the results of a global analysis of the 2HDM$I\!I$ parameter space using all relevant theoretical and experimental constraints in a fit performed with the CKMfitter package. The findings of our publication \cite{Baglio:2014nea} did not change by adding CMS exclusion limits on $H\to hh$ and $A\to hZ$ decays: At $95$\% CL, we can set a lower limit of $240$ GeV to $m_A$, while $m_H$ can be smaller than $200$ GeV. The triple Higgs coupling $g_{hhh}$ cannot exceed its SM value $g_{hhh}^{SM}$, whereas the coupling $g_{hhH}$ can be more than twice as large as $g_{hhh}^{SM}$. Both couplings are important for the future measurement of the $gg\to hh$ cross section, where we showed for a chosen set of benchmark scenarios that enhancements of the expected SM value by almost a factor of $70$ are still possible in the 2HDM$I\!I$. Further sets of benchmark points as well as results for triple Higgs couplings and $H$ and $A$ branching fractions can be found in our latest article \cite{Baglio:2014nea}.

\section*{Acknowledgments}

These proceedings are based on results obtained in collaboration with J. Baglio, U. Nierste and M. Wiebusch, whom I also thank for proofreading. Moreover, I want to thank the organizers of the electroweak session of the ``Rencontres de Moriond 2014'' for a pleasant week full of a very interesting and broad programme. The research leading to these results has received funding from the European Research Council under the European Union's Seventh Framework Programme (FP/2007-2013) / ERC Grant Agreement n.~279972.

\section*{References}

\bibliography{Proceedings_Moriond2014_Eberhardt}

\end{document}